\documentclass[]{spie}  

\usepackage[]{graphicx}

\title{Quantum phase uncertainty in mutually unbiased measurements
 and Gauss sums
}

\author{Michel Planat\supit{a} and Haret Rosu\supit{b}
\skiplinehalf \supit{a}Institut FEMTO-ST, Departement LPMO,\\ 32
Avenue de
l'Observatoire, 25044 Besan\c{c}on Cedex, France; \\
\supit{b}Div. of Advanced Materials, \\IPICyT, Apdo Postal 3-74,
Tangamanga, San Luis Potos\'{i}, Mexico.  }

\authorinfo{Further author information: (Send correspondence to Michel Planat)\\Michel Planat: E-mail: planat@lpmo.edu, Telephone: 33 3 81 85 39 57}

\begin{document}
  \maketitle

\begin{abstract}
Mutually unbiased bases (MUBs), which are such that the inner
product between two vectors in different orthogonal bases is
constant equal to the inverse $1/\sqrt{d}$, with $d$ the dimension
of the finite Hilbert space, are becoming more and more studied
for applications such as quantum tomography and cryptography, and
in relation to entangled states and to the Heisenberg-Weil group
of quantum optics. Complete sets of MUBs of cardinality $d+1$ have
been derived for prime power dimensions $d=p^m$ using the tools of
abstract algebra (Wootters in 1989, Klappenecker in 2003).
Presumably, for non prime dimensions the cardinality is much less.

The bases can be reinterpreted as quantum phase states, i.e. as
eigenvectors of Hermitean phase operators generalizing those
introduced by Pegg \& Barnett in 1989. The MUB states are related
to additive characters of Galois fields (in odd characteristic p)
and of Galois rings (in characteristic 2). Quantum Fourier
transforms of the components in vectors of the bases define a more
general class of MUBs with multiplicative characters and additive
ones altogether. We investigate the complementary properties of
the above phase operator with respect to the number operator. We
also study the phase probability distribution and variance for
physical states and find them related to the Gauss sums, which are
sums over all elements of the field (or of the ring) of the
product of multiplicative and additive characters.

Finally we relate the concepts of mutual unbiasedness and maximal
entanglement. This allows to use well studied algebraic concepts
as efficient tools in our quest of minimal uncertainty in quantum
information primitives.

\end{abstract}



\section{INTRODUCTION}
\label{sect:intro}

In quantum mechanics, orthogonal bases of a Hilbert space
$\mathcal{H}_q$ of finite dimension $q$ are mutually unbiased if
inner products between all possible pairs of vectors of distinct
bases equal $1/\sqrt{q}$. They are also said to be maximally non
commutative in the sense that a measurement over one basis leaves
one completely uncertain as to the outcome of a measurement
performed over a basis unbiased to the first. Eigenvectors of
ordinary Pauli spin matrices (i.e. in dimension $q=2$) provide the
best known example. With a complete set of $q+1$ mutually unbiased
measurements one can ascertain the density matrix of en ensemble
of unknown quantum $q$-states, so that a natural question emerges
as which mathematics may provide the construction. It will be
shown that in dimension $q=p^m$ which is the power of a prime $p$,
the complete sets of mutually unbiased bases (MUBs) result from
Fourier analysis over a Galois field $F_q$ (in odd characteristic
$p$)\cite{Wootters89} or of Galois ring $R_{4^m}$(in even
characteristic $2$)\cite{Klapp03} . An exhaustive literature on
MUBs can be found in\cite{Wootters04bis} , \cite{Planat04} .
Complete sets of MUBs have an intrinsic geometrical
interpretation, and were related to discrete phase
spaces\cite{Wootters04bis} , \cite{Paz} , \cite{Pittinger05} ,
finite projective planes\cite{Saniga} , \cite{Saniga2} , convex
polytopes \cite{Bengtson} , and complex projective $2$-designs
\cite{Barnum02} , \cite{Klap2} . The last paper points out the
relation to symmetric informationally complete positive operator
measures (SIC-POVMs)\cite{Wootters04} , \cite{Grassl04} ,
\cite{Appleby04} and to Latin squares\cite{Wocjan04} .

A Galois field is a finite set structure endowed with two group
operations, the addition \lq\lq$+$" and the multiplication
\lq\lq$\cdot$". The field $F_q$ can be represented as classes of
polynomials obtained by computing modulo an irreducible polynomial
over the ground field $F_p=\mathcal{Z}_p$, the integers modulo
$p$\cite{Lidl83}. A Galois field exists if and only if (iff)
$q=p^m$.

A character $\kappa(g)$ over an abelian group $G$ is a
(continuous) map from $G$ to the field of complex numbers
$\mathcal{C}$, which is of modulus $1$, i.e. such that
$|\kappa(g)|=1$, $g\in G$. The multiplicative characters
$\psi_k(n)=\exp(\frac{2i\pi n k}{q})$, $k=0..q-1$ are well known
since they constitute the basis for the ordinary discrete Fourier
transform. But the additive characters introduced below are the
ones which are useful to construct the MUBs. This construction is
implicit in some previous
papers\cite{Wootters89},\cite{Klapp03},\cite{Planat04}, and is now
being fully recognized\cite{Durt04} , \cite{Klimov04} .

An interesting consequence is as follows: the discrete Fourier
transform in $\mathcal{Z}_q$ has been used as a definition of
phase states $|\theta_k \rangle$, $k=0..q-1$ in $\mathcal{H}_q$.
The phase states \cite{Pegg89} could be considered as eigenvectors
of a properly defined Hermitian phase operator $\Theta$. Phase
properties and phase fluctuations attached to particular field
states were extensively described. In particular the classical
phase variance $\pi^2/3$ could be recovered. Similarly a phase
operator $\Theta_{\rm{Gal}}$ having the MUBs as eigenvectors will
be constructed here, but in contrast to the case of $\Theta$,
phase fluctuations from $\Theta_{\rm{Gal}}$ can in principle be
reduced, a result which reflects the property of Gauss sums over
$F_q$, and which confirms the interest of MUBs for quantum signal
processing. Character sums and Gauss sums which are useful for
optimal bases of $m$-qudits ($p$ odd) will be also generalized to
optimal bases of $m$-qubits ($p=2$).

It is worthwhile to mention that quadratic Gauss sums were already
met in the transient and revival dynamics of semi-classical wave
packets \cite{Berry01} . Finally, related exponential sums:
Ramanujan sums and Kloosterman sums were found to control the
phase dynamics of quantum phase-locked states \cite{Planat03} .

\section{Some character sums over a Galois field}
\label{sect:Chars}

Let us consider the field of polynomials $F_p[x]$ defined over the
field $F_p$
\begin{equation}
F_p[x]=\{a_0+a_1 x+\cdots+a_n x^n\},~~a_i\in F_p.
\end{equation}
For a polynomial $g\in F_p[x]$, the residue class ring
$F_p[x]/(g)$, where $(g)$ is the ideal class generated by $g$ is a
field iff g is irreducible over $F_p$ (it cannot be factored over
$F_p$).

For example for $q=2^2$, one can choose the polynomial
$g(x)=x^2+x+1 \in F_2[x]$ which is irreducible over $F_2$.
Contrary to $\mathcal{Z}_4$ which has zero divisors and is thus
only a ring, the above construction defines the field with four
residue classes: $F_4=\{0,1,x,x+1\}$.

The key relation between Galois fields $F_q$ and mutually unbiased
bases is the theory of characters. It starts from a map from the
extended field $F_q$ to the ground field $F_p$ which is called the
trace function
\begin{equation}
tr(x)=x+x^p+\cdots+x^{p^{m-1}}\in F_p,~~ \forall ~x\in F_q.
\label{trace0}
\end{equation}
In addition to its property of mapping an element of $F_q$ into
$F_p$, the trace function has the properties
\begin{eqnarray}
&tr(x+y)=tr(x)+ tr(y),~~x,y \in F_q\nonumber \\
&tr(ax)=atr(x),~~x \in F_q,~a \in F_p,\nonumber \\
&tr(a)=ma, ~~a\in F_p,\nonumber\\
&tr(x^q)=tr(x),~~ x\in F_q. \label{trace}
\end{eqnarray}
Using (\ref{trace0}), an additive character over $F_q$ is defined
as
\begin{equation}
\kappa(x)=\omega_p^{tr(x)},~~\omega_p=\exp(\frac{2i\pi}{p}),~~x
\in F_q.
\end{equation}
It satisfies $\kappa(x+y)=\kappa(x)\kappa(y), ~x,y \in F_q$.

The multiplicative characters are
\begin{equation}
\psi_k(n)=\exp(\frac{2i\pi n k}{q}),~~ k=0..q-1,~~ k=0..q-1.
\end{equation}
The construction of MUBs will be related to character sums with
polynomial arguments $f(x)$ also called Weil sums\cite{Klapp03}
\begin{equation}
\sum_{x \in F_q}\kappa(f(x)). \label{Weilsums}
\end{equation}
In particular ( theorem 5.38 in \cite{Lidl83}), for a polynomial
$f(x) \in F_q[x]$ of degree $d\ge 1$, with $gcd(d,q)=1$, one gets
$|\sum_{x \in F_q}\kappa(f(x))|\le(d-1)q^{1/2}$.

Finally, the phase fluctuations arising from MUBs (quantum phase
states) will be found to be related to the Gauss sums of the form
\begin{equation}
G(\psi,\kappa)=\sum_{x \in F_q^*}\psi(x)\kappa(x). \label{Gauss}
\end{equation}
Using the notation $\psi_0$ for a trivial multiplicative character
$\psi=1$, and $\kappa_0$ for a trivial additive character
$\kappa=1$ the Gaussian sums (\ref{Gauss}) satisfy
$G(\psi_0,\kappa_0)=q-1$; $G(\psi_0,\kappa)=-1$;
$G(\psi,\kappa_0)=0$ and $|G(\psi,\kappa)|=q^{1/2}$ for nontrivial
characters $\kappa$ and $\psi$.

We also mention that more general Gauss sums were studied as
\begin{equation}
G(\psi,\kappa)=\sum_{x \in F_q}\psi(f(x))\kappa(g(x)),
\label{Gaussnew}
\end{equation}
with $f,g \in F_q[x]$ and found to be of the order of magnitude
$\sqrt{q}$ (\cite{Lidl83} p. 249).

\section{Mutually unbiased bases of quantum phase states \\(in odd prime characteristic)}
\label{sect:MUBs}

Let us introduce a class of quantum states as a \lq\lq Galois"
Fourier transform
\begin{equation}
|\theta^{(y)}\rangle=\frac{1}{\sqrt{q}}\sum_{n \in
F_q}\psi(n)\kappa(y n)|n\rangle,~~y \in F_q
 \label{MUB}
\end{equation}
in which the coefficient in the computational base
$\{|0\rangle,|1\rangle,\cdots,|q-1\rangle\}$ represents the
product of an arbitrary multiplicative character $\psi_k(n)$ by an
arbitrary additive character $\kappa(yn)$.

\subsection{Pegg \& Barnett 89}

For $\kappa=\kappa_0$ and $\psi\equiv \psi_k(n)$, one recovers the
ordinary quantum Fourier transform over $\mathcal{Z}_q$. It has
been shown\cite{Pegg89} that the corresponding states
\begin{equation}
|\theta_k\rangle=\frac{1}{\sqrt{q}}\sum_{n \in
\mathcal{Z}_q}\psi_k(n)|n\rangle, \label{Pegg}
\end{equation}
are eigenstates of the Hermitian phase operator
\begin{equation}
\Theta=\sum_{k\in \mathcal{Z}_q}\theta_k|\theta_k\rangle \langle
\theta_k|,
\end{equation}
with eigenvalues $\theta_k=\theta_o+\frac{2\pi k}{q}$, $\theta_0$
an arbitrary initial phase.

\subsection{Wootters \& Fields 89}

We employ the Euclidean division theorem (theorem 11.19 in
\cite{Lidl98}) for the field $F_q$, which says that given any two
polynomials $y$ and $n$ in $F_q$, there exists a uniquely
determined pair $(a,b)\in F_q \times F_q$, such that $y=an+b$,
$\deg(b)<deg(a)$. Using the decomposition of the exponent in
(\ref{MUB}), we obtain
\begin{equation}
|\theta_b^a\rangle=\frac{1}{\sqrt{q}}\sum_{n\in
F_q}\psi_k(n)\kappa(an^2+bn)|n\rangle,~~a,b \in F_q. \label{MUB1}
\end{equation}
This defines a set of $q$ bases (with index $a$) of $q$ vectors
(with index $b$). Using Weil sums (\ref{Weilsums}) it is easily
shown that, for $q$ odd, so that $\rm{gcd}(2,q)>1$, the bases are
orthogonal and mutually unbiased to each other and to the
computational base.
More precisely
\begin{equation}
|\langle \theta_b^a|\theta_d^c\rangle|=|\frac{1}{q}\sum_{n\in F_q}
\omega_p^{tr((c-a)n^2+(d-b)n}|=\left\{\begin{array}{ll}
&\delta_{bd}~~\mbox{if}~c=a~(\mbox{orthogonality}) \\
&\frac{1}{\sqrt{q}} ~~\mbox{if}~c\neq a~(\mbox{unbiasedness}).
\end{array}\right.
\end{equation}
The MUB states are also eigenstates of a \lq\lq Galois" quantum
phase operator
\begin{equation}
\Theta_{\rm{Gal}}=\sum_{b\in F_q}\theta_b|\theta_b^a\rangle
\langle \theta_b^a|,~~a,b \in F_q. \label{GalMUB}
\end{equation}
with eigenvalues  $\theta_b=\frac{2\pi b}{q}$.

\section{Quantum phase fluctuations from MUBs \\(in odd prime  characteristic)}
\label{MUBodd}

\subsection{The Galois operator}

Using (\ref{MUB1}) in (\ref{GalMUB}) and the properties of the
field theoretical trace (\ref{trace}), the Galois operator reads
\begin{equation}
\Theta_{\rm{Gal}}=\frac{2\pi}{q^2}\sum_{m,n \in
F_q}\psi_k(n-m)\omega_p^{tr[a(n^2-m^2)]} S(n,m) |n\rangle \langle
m|,~~\rm{with}~S(n,m)=\sum_{b\in F_q}b\omega_p^{tr[b(n-m)]}.
\label{GalMUB1}
\end{equation}
In the diagonal matrix elements, we have the partial sums
\begin{equation}
S(n,n)=\frac{q(q-1)}{2}, \label{Snn}
\end{equation}
so that $\langle n|\Theta_{\rm{Gal}}|n\rangle=\frac{\pi
(q-1)}{q}$. In the non-diagonal matrix elements, the partial sums
can be calculated from
\begin{equation}
\sum_{b\in F_q}b
x^b=x(1+2x+3x^2+\cdots+qx^{q-1})=x[\frac{1-x^q}{(1-x)^2}-\frac{qx^q}{1-x}]=\frac{xq}{x-1},
\end{equation}
where we introduced $x=\omega_p^{tr(n-m)}$ and we made use of the
relation $x^q=1$. Finally
\begin{equation}
S(m,n)=\frac{q}{1-\omega_p^{tr(m-n)}}. \label{Smn_sums}
\end{equation}

\subsection{The Galois phase-number commutator}

Using (\ref{GalMUB1}) and the number operator
\begin{equation}
N=\sum_{l\in F_q}l |l\rangle \langle l|,
\end{equation}
the matrix elements of the phase-number commutator
$[\Theta_{\rm{Gal}},N]$ are calculated as
\begin{equation}
u_{\rm{Gal}}(n,m)=\frac{2
\pi}{q^2}(n-m)\psi_k(n-m)\omega_p^{tr[a(n^2-m^2)]}S(n,m).
\end{equation}
The diagonal elements vanish, the corresponding matrix is
antihermitian since
$u_{\rm{Gal}}(n,m)=-u_{\rm{Gal}}^{\dagger}(m,n)$, and the states
are pseudo-classical since $\lim_{q \rightarrow
\infty}u_{\rm{Gal}}(n,m)=0$.

\subsection{Galois phase expectation value and variance}

For the evaluation of phase properties of MUB states we consider a
pure phase state of the form
\begin{equation}
|f\rangle=\sum_{n\in
F_q}u_n|n\rangle,~~\rm{with}~u_n=\frac{1}{\sqrt{q}}\exp(in \beta),
\end{equation}
where $\beta$ is a real parameter, and we compute respectively the
phase probability distribution, the phase expectation value and
the phase variance as
\begin{eqnarray}
&|<\theta_b|f>|^2, \nonumber \\
&  <\Theta_{\rm{Gal}}>=\sum_{b \in F_q}\theta_b|<\theta_b|f>|^2,
\nonumber \\
&<\Delta\Theta_{\rm{Gal}}^2>=\sum_{b \in
F_q}(\theta_b-<\Theta_{\rm{Gal}}>)^2|<\theta_b|f>|^2,
\end{eqnarray}
where the upper index $a$ for the base is implicit and we omitted
it for simplicity.

\subsection{Phase expectation value}

The two factors in the expression for the probability distribution
\begin{equation}
 \frac{1}{q^2} [\sum_{n \in F_q} \psi_k(-n) \exp(i n \beta)\omega_p^{-tr(an^2+bn)}][\sum_{m \in F_q} \psi_k(m) \exp(-i m
\beta)\omega_p^{tr(am^2+bm)}],
\end{equation}
have absolute values bounded by the absolute value of generalized
Gauss sums (\ref{Gaussnew}), so that the overall bound is
\begin{equation}
|<\theta_b|f>|^2\le \frac{1}{q}.
\end{equation}
It follows that the absolute value of the phase expectation value
is bounded as it is expected for an arbitrary phase factor.
\begin{equation}
|<\Theta_{\rm{Gal}}>|\le \frac{2\pi}{q^2}\sum_{b \in F_q}b\le \pi.
\end{equation}
More precisely  the phase expectation value can be expressed as
\begin{equation}
<\Theta_{\rm{Gal}}>=\frac{2 \pi}{q^3}\sum_{m,n \in
F_q}\psi_k(m-n)\exp[i(n-m)\beta]\omega_p^{tr[a(m^2-n^2]}S(m,n).
\label{expect}
\end{equation}
where the sums $S(m,n)$ were defined in (\ref{Snn}) and
(\ref{Smn_sums}). All the $q$ diagonal terms $m=n$ in
$<\Theta_{\rm{Gal}}>$ contribute an order of magnitude
$\frac{2\pi}{q^3}q S(n,n)\simeq \pi$. The contribution of
off-diagonal terms and possible cancellation of phase oscillations
could be considered from numerical plots, since the sums in
(\ref{expect}) are not easy to evaluate analytically.

\subsection{Phase variance}

The phase variance can be written as
\begin{equation}
<\Delta\Theta_{\rm{Gal}}^2>=\sum_{b \in F_q}(\theta_b^2-2 \theta_b
<\Theta_{\rm{Gal}}>)|<\theta_b|f>|^2,
 \label{variance}
\end{equation}
The coefficient $<\Theta_{\rm{Gal}}>^2\sum_{b\in
F_q}|<\theta_b|f>|^2$ doesn't contribute since it is proportional
to the Weil sum $\sum_{b \in F_q}\omega_p^{tr(b(n-m)}=0$. As a
result a cancellation of phase fluctuations may occur in
(\ref{variance}) from to the two extra terms of opposite sign.

But the calculation are again not easy to perform analytically.
For the first term one gets
\begin{equation}
\frac{4 \pi^2}{q^4} \sum_{m,n \in
F_q}\psi_k(m-n)\exp[i(n-m)\beta]\omega_p^{tr[a(m^2-n^2)]}T(m,n)
,~~\rm{with}~T(n,m)=\sum_{b\in F_q}b^2\omega_p^{tr[b(n-m)]}.
\label{FirstTerm}
\end{equation}
In the diagonal elements we have the partial sums
\begin{equation}
T(n,n)=\sum_{b\in F_q}b^2=\frac{q^3}{3}-\frac{q^2}{2}+\frac{q}{6}.
\end{equation}

In the non-diagonal terms the partial sums can be calculated from
\begin{equation}
\sum_{b\in F_q}b^2 x^b=x(1+2^2x+3^2x^2+\cdots+(q-1)^2x^{q-1})=x
\frac{d}{dx}\{
x[\frac{1-x^q}{(1-x)^2}-\frac{qx^q}{1-x}]\}=\frac{-2qx}{(1-x)^2},
\end{equation}
where we introduced $x=\omega_p^{tr(n-m)}$ and we made use of the
relations $x^q=1$ and $q^2=0$.

The second term in (\ref{variance}) is
\begin{equation}
\sum_{b \in F_q}-2 \theta_b
<\Theta_{\rm{Gal}}>|<\theta_b|f>|^2=-2<\Theta_{\rm{Gal}}>^2.
\label{SecondTerm}
\end{equation}
Partial cancellation occurs in diagonal terms of (\ref{variance})
since the contribution is
\begin{equation}
\frac{4\pi^2}{q^4}q T(n,n)-\frac{8 \pi^2}{q^4}S(n,n)^2\simeq
\frac{4\pi^2}{3}-2\pi^2=-\frac{2\pi^2}{3},
\end{equation}
which is still twice (in absolute value) the amount of phase
fluctuations in the classical regime. It is expected that
cancellations also occur in the non-diagonal terms to beat the
classical limit, as for the case of squeezed states.

\section{Galois rings and their character sums}

\subsection{Construction of the Galois rings of characteristic $4$}

 The Weil sums (\ref{Weilsums}) which have been proved useful
in the construction of MUB's in odd characteristic $p$ (and odd
dimension $q=p^m$), are not useful in characteristic $p=2$, since
in this case the degree $2$ of polynomial $f(x)$ is such that
$\rm{gcd}(2,q)$=0.

An elegant method for constructing complete sets of MUBs of
$m$-qubits was found  \cite{Klapp03} . It makes use of objects of
the context of quaternary codes \cite{Wan97} , the so-called
Galois rings $R_{4^m}$. In contrast to the Galois fields where the
ground alphabet has $p$ elements ($p$ a prime number) in the field
$F_p=\mathcal{Z}_p$, the ring $R_{4^m}$ takes its ground alphabet
in $\mathcal{Z}_4$. To construct it one uses the ideal class
$(h)$, where $h$ is a (monic) basic irreducible polynomial of
degree $m$. It is such that its restriction to $\bar{h}(x)=h(x)
\rm{mod}~ 2$ is irreducible over $\mathcal{Z}_2$. The Galois ring
$R_{4^m}$ is defined as the residue class ring
$\mathcal{Z}_4[x]/(h)$. It has cardinality $4^m$.

We also needs the concept of a primitive polynomial. A (monic)
primitive polynomial, of degree $m$, in the field $F_q[x]$ is
irreducible over $F_q$ and has a root $\alpha \in F_{q^m}$ that
generates the multiplicative group of $F_{q^m}$. A polynomial
$f\in F_q[x]$ of degree $m$ is primitive iff $f(0)\neq 0$ and
divides $x^r-1$, with $r=q^m-1$.

Similarly for Galois rings $R_{4^m}$, if $\bar{h}[x]$ is a
primitive polynomial of degree $m$ in $\mathcal{Z}_2[x]$, then
there is a unique basic primitive polynomial $h(x)$ of degree $m$
in $\mathcal{Z}_4[x]$ (it divides $x^r-1$, with $r=2^m-1$). It can
be found as follows \cite{Hammons94}. Let $\bar{h}(x)=e(x)-d(x)$,
where $e(x)$ contains only even powers and $d(x)$ only odd powers;
then $h(x^2)=\pm(e^2(x)-d^2(x))$. For $m=2$, $3$ and $4$ one takes
$\bar{h}(x)=x^2+x+1$, $\bar{h}(x)=x^3+x+1$ and
$\bar{h}(x)=x^4+x+1$ and one gets $h(x)=x^2+x+1$, $x^3+2x^2+x-1$
and $x^4+2x^2-x+1$, respectively.

Any non zero element of $F_{p^m}$ can be expressed in terms of a
single primitive element. This is no longer true in $R_{4^m}$,
which contains zero divisors. But in the latter case there exists
a nonzero element $\xi$ of order $2^m-1$ which is a root of the
basic primitive polynomial $h(x)$. Any element $y \in R_{4^m}$ can
be uniquely determined in the form $y=a + 2 b$, where $a$ and $b$
belong to the so-called Teichm\"{u}ller set $\mathcal{T}_m =
(0,1,\xi,\cdots,\xi^{2^m-2})$. Moreover, one finds that
$a=y^{2^m}$. We can also define the trace to the base ring
$\mathcal{Z}_4$ by the map
\begin{equation}
\tilde{tr}(y)=\sum_{k=0}^{m-1}\sigma^k(y), \label{trace2}
\end{equation}
where the summation runs over $R_{4^m}$ and the Frobenius
automorphism $\sigma$ reads
\begin{equation}
\sigma(a+2 b)=a^2+ 2 b^2.
\end{equation}
Let us apply this formula to the case $m=2$ (which will correspond
to 2-qubits). In $R_{4^2}=\mathcal{Z}_4[x]/(x^2+x+1)$ the
Teichm\"{u}ller set reads $\mathcal{T}_2=(0,1,x,3+3x)$; the $16$
elements $a+2 b$ with $a$ and $b$ in $\mathcal{T}_2$ are shown in
the following matrix
\begin{equation}
 \left [\begin{array}{cccc} 0 & 2&2x&2+2x\\ 1 & 3&1+2x&3+2x\\
 x&2+x&3x&2+3x\\3+3x&1+3x&3+x&1+x\nonumber \\
\end{array}\right].
\end{equation}
The case $m=3$ corresponding to $3$-qubits can be examined in a
similar fashion, with the ring
$R_{4^3}=\mathcal{Z}_4[x]/(x^3+2x^2+x-1)$ and the Teichm\"{u}ller
set featuring the following eight elements:
$\mathcal{T}_3=\{0,1,x,x^2,1+3x+2x^2,2+3x+3x^2,3+3x+x^2,1+2x+x^2\}$.

In the Galois ring of characteristic $4$ the additive characters
are
\begin{equation}
\tilde{\kappa}(x)=\omega_4^{\tilde{\rm{tr}}(x)}=i^{\tilde{\rm{tr}}(x)}.
\end{equation}
\subsection{Exponential sums over $R_{4^m}$}

The Weil sums (\ref{Weilsums}) are replaced by the exponential
sums \cite{Klapp03}
\begin{equation}
\Gamma(y)=\sum_{u \in \mathcal{T}_m}\tilde{\kappa}(y u),~~y \in
R_{4^m} \label{newWeilsums}
\end{equation}
which satisfy
\begin{equation}
|\Gamma(y)|=\left\{\begin{array}{ll}
&0~~\mbox{if}~y \in 2 \mathcal{T}_m,~y \ne 0 \\
&2^m ~~\mbox{if}~y=0\\
&\sqrt{2^m}~~\mbox{otherwise}.
\end{array}\right.
\end{equation}
Gauss sums for Galois rings were constructed \cite{Oh01}
\begin{equation}
G_y(\tilde{\psi},\tilde{\kappa})=\sum_{x \in R_{4^m}}
\tilde{\psi}(x)\tilde{\kappa}(yx), ~~y \in
R_{4^m},\label{GaussGal}
\end{equation}
where the multiplicative character $\bar{\psi}(x)$ can be made
explicit \cite{Oh01} .

Using the notation $\bar{\psi_0}$ for a trivial multiplicative
character and $\tilde{\kappa_0}$ for a trivial additive character,
the Gaussian sums (\ref{GaussGal}) satisfy
$G(\tilde{\psi_0},\tilde{\kappa_0})=4^m$;
$G(\tilde{\psi},\tilde{\kappa_0})=0$ and
$|G(\tilde{\psi},\tilde{\kappa})|\le 2^m$.

\section{Mutually unbiased bases of quantum phase states\\ ($m$-qubits)}
\label{sect:MUB2s}

The quantum phase states for $m$-qubits can be found as the \lq\lq
Galois ring" Fourier transform

\begin{equation}
|\theta^{(y)}\rangle=\frac{1}{\sqrt{2^m}}\sum_{n \in
\mathcal{T}_m}\tilde{\psi}(n)\tilde{\kappa}(y n)|n\rangle,~~y \in
R_{4^m}.
 \label{MUBpair}
\end{equation}

\subsection{A. Klappenecker \& M. R\"{o}tteler 03}

It was shown in the previous section that each element $y$ of the
ring $R_{4^m}$ decomposes as $y=a+2b$, $a$ and $b$ in the
Teichm\"{u}ller set $\mathcal{T}_m$. Using this result in the
character function $\tilde{\kappa}$ one obtains
\begin{equation}
|\theta_b^a\rangle=\frac{1}{\sqrt{2^m}}\sum_{n\in
\mathcal{T}_m}\tilde{\psi}_k(n)\tilde{\kappa}[(a+2b)n]|n\rangle,~~a,b
\in \mathcal{T}_m. \label{MUB1pair}
\end{equation}
This defines a set of $2^m$ bases (with index $a$) of $2^m$
vectors (with index $b$). Using the exponential sums
(\ref{newWeilsums}), it is easy to show that the bases are
orthogonal and mutually unbiased to each other and to the
computational base. The case $\bar{\psi}\equiv \bar{\psi_0}$ was
obtained before \cite{Klapp03} .

\subsection{MUB's for m-qubits, $m=1$, $2$ and $3$}

For the special case of qubits, one uses $\tilde{\rm{tr}}(x)=x$ in
(\ref{MUB1pair}) so that the three pairs of MUB's are given as
\begin{equation}
[|0\rangle,|1\rangle];~~\frac{1}{\sqrt{2}}[|0\rangle+|1\rangle,|0\rangle-|1\rangle];~~\frac{1}{\sqrt{2}}[|0\rangle+
i |1\rangle,|0\rangle- i |1\rangle]. \nonumber
\end{equation}

For $2$-qubits one gets a complete set of $5$ bases
\begin{eqnarray}
&(|0\rangle,|1\rangle,|2\rangle,|3\rangle); \nonumber\\
&\frac{1}{2}[|0\rangle+|1\rangle+|2\rangle+|3\rangle,|0\rangle+|1\rangle-|2\rangle-|3\rangle,
 |0\rangle-|1\rangle-|2\rangle+|3\rangle,|0\rangle-|1\rangle+|2\rangle-|3\rangle]\nonumber\\
&\frac{1}{2}[|0\rangle-|1\rangle-i|2\rangle-i|3\rangle,|0\rangle-|1\rangle+i|2\rangle+i|3\rangle,
 |0\rangle+|1\rangle+i|2\rangle-i|3\rangle,|0\rangle+|1\rangle-i|2\rangle+i|3\rangle]\nonumber\\
 &\frac{1}{2}[|0\rangle-i|1\rangle-i|2\rangle-|3\rangle,|0\rangle-i|1\rangle+i|2\rangle+|3\rangle,
 |0\rangle+i|1\rangle+i|2\rangle-|3\rangle,|0\rangle+i|1\rangle-i|2\rangle+|3\rangle]\nonumber\\
 &\frac{1}{2}[|0\rangle-i|1\rangle-|2\rangle-i|3\rangle,|0\rangle-i|1\rangle+|2\rangle+i|3\rangle,
 |0\rangle+i|1\rangle+|2\rangle-i|3\rangle,|0\rangle+i|1\rangle-|2\rangle+i|3\rangle],\nonumber\\
 \label{quartrits}
\end{eqnarray}

and for $3$ qubits a complete set of $9$ bases

\begin{eqnarray}
&(|0\rangle,|1\rangle,|2\rangle,|3\rangle,|4\rangle,|5\rangle,|6\rangle,|7\rangle); \nonumber\\
&\frac{1}{4}[|0\rangle+|1\rangle+|2\rangle+|3\rangle+|4\rangle+|5\rangle+|6\rangle+|7\rangle,
|0\rangle+|1\rangle-|2\rangle+|3\rangle-|4\rangle-|5\rangle-|6\rangle+|7\rangle,\nonumber\\
&|0\rangle-|1\rangle+|2\rangle-|3\rangle-|4\rangle-|5\rangle+|6\rangle-|7\rangle,
|0\rangle+|1\rangle-|2\rangle-|3\rangle-|4\rangle+|5\rangle+|6\rangle-|7\rangle,\nonumber\\
&|0\rangle-|1\rangle-|2\rangle-|3\rangle+|4\rangle+|5\rangle-|6\rangle+|7\rangle,
|0\rangle-|1\rangle-|2\rangle+|3\rangle+|4\rangle-|5\rangle+|6\rangle-|7\rangle,\nonumber\\
&|0\rangle-|1\rangle+|2\rangle+|3\rangle-|4\rangle+|5\rangle-|6\rangle-|7\rangle,
|0\rangle+|1\rangle+|2\rangle-|3\rangle+|4\rangle-|5\rangle-|6\rangle-|7\rangle],\nonumber\\
&\cdots
\end{eqnarray}
where only the first two bases have been printed for simplicity.

\subsection{Quantum phase fluctuations for $m$-qubits}
Quantum phase states of $m$-qubits (\ref{MUB1pair}) derive from a
\lq\lq Galois ring" quantum phase operator as in (\ref{GalMUB}),
and calculations similar to those performed in Sect.
(\ref{MUBodd}) can be done, since the $\tilde{\rm{tr}}$ operator
(\ref{trace2}) follows rules similar to the $\rm{tr}$ operator
(\ref{trace0}). In analogy to the case of qdits in dimension
$p^m$, $p$ an odd prime, phase properties for sets of $m$-qubits
heavily rely on the Gauss sums (\ref{GaussGal}). As before the
calculations are tedious but can in principle be achieved in
specific cases.

\section{Mutual unbiasedness and maximal entanglement}

It has been shown in this paper that there is a founding link
between irreducible polynomials over a ground field $F_p$ and
complete sets of mutually unbiased bases arising from Fourier
transform over a lifted field $F_q$, $q=p^m$, $p$ a prime number.
On the other hand the physical concept of entanglement over the
Hilbert space $\mathcal{H}_q$ evokes irreducibility. Roughly
speaking entangled states in $\mathcal{H}_q$ cannot be factored
into tensorial products of states in Hilbert spaces of lower
dimension. We show now that there is an intrinsic relation between
MUBs and maximal entanglement.

We are familiar we the Bell states
\begin{eqnarray}
&(|\mathcal{B}_{0,0}\rangle,|\mathcal{B}_{0,1}\rangle)=\frac{1}{\sqrt{2}}(|00\rangle+|11\rangle,|00\rangle-|11\rangle),\nonumber\\
&(|\mathcal{B}_{1,0}\rangle,|\mathcal{B}_{1,1}\rangle)=\frac{1}{\sqrt{2}}(|01\rangle+|10\rangle,|01\rangle-|10\rangle),
\nonumber
\end{eqnarray}
where a compact notation $|00\rangle=|0\rangle\odot|0\rangle$,
$|01\rangle=|0\rangle\odot|1\rangle$,\dots, is employed for the
tensorial products.

These states are both orthonormal and maximally entangled, such
that $trace_2|\mathcal{B}_{h,k}\rangle \langle\mathcal{B}_{h,k}|
=\frac{1}{2}I_2$, where $trace_2$ means the partial trace over the
second qubit \cite{Nielsen00} .

One can define more generalized Bell states using the
multiplicative Fourier transform (\ref{Pegg}) applied to the
tensorial products of two qudits \cite{Cerf01} , \cite{Durt04}
\begin{equation}
|\mathcal{B}_{h,k}\rangle=\frac{1}{\sqrt{q}}\sum_{n=0}^{q-1}\omega_q^{k
n}|n,n+h\rangle, \label{FourierEntang}
\end{equation}
These states are both orthonormal, $\langle
\mathcal{B}_{h,k}|\mathcal{B}_{h',k'} \rangle
=\delta_{hh'}\delta_{kk'}$, and maximally entangled,
$trace_2|\mathcal{B}_{h,k}\rangle \langle\mathcal{B}_{h,k}|
=\frac{1}{q}I_q$.

But we can also define a more general class of maximally entangled
states using the Fourier transform over $F_q$ (\ref{MUB1}) as
follows
\begin{equation}
|\mathcal{B}_{h,b}^a\rangle=\frac{1}{\sqrt{q}}\sum_{n=0}^{q-1}\omega_p^{tr[(a
 n + b)n ]}|n,n+ h\rangle,
\label{entangledGalois}
\end{equation}
A list of the generalized Bell states of qutrits for the base
$a=0$ can be found in \cite{Fujii01} , the work that relies on a
coherent state formulation of entanglement. In general, for $q$ a
power of a prime, starting from (\ref{entangledGalois}) one
obtains $q^2$ bases of $q$ maximally entangled states. Each set of
the $q$ bases (with $h$ fixed) has the property of mutual
unbiasedness.

Similarly for sets of maximally entangled m-qubits one uses the
Fourier transform over Galois rings (\ref{MUB1pair}) so that
\begin{equation}
|\mathcal{B}_{h,b}^a\rangle=\frac{1}{\sqrt{2^m}}\sum_{n=0}^{2^m-1}i^{tr[(a
+  2 b) n]}|n,n+ h\rangle \label{twoevendits}.
\end{equation}
For qubits ($m=1$) one gets the following bases of maximally
entangled states (in matrix form, safe for the proportionality
factor)
\begin{equation}
 \left [\begin{array}{cc} (|00\rangle+|11\rangle,|00\rangle-|11\rangle);&(|01\rangle+|10\rangle,|01\rangle-|10\rangle)
 \\(|00\rangle+i|11\rangle,|00\rangle-i|11\rangle);&(|01\rangle+i|10\rangle,|01\rangle)-i|10\rangle)\\
\end{array}\right].
\end{equation}
Two bases in one column are mutually unbiased, while vectors in
two bases on the same line are orthogonal to each other.

For $2$-particle sets of quartits, using Eqs.(\ref{quartrits}) and
(\ref{twoevendits}), one gets $4$ sets
$(|\mathcal{B}_{h,b}^a\rangle$, $ h=0,...,3)$ of $4$ MUBs
$(a=0,...,3)$
\begin{eqnarray}
&\{(|00\rangle+|11\rangle+|22\rangle+|33\rangle,|00\rangle+|11\rangle-|22\rangle-|33\rangle,\nonumber
\\
&|00\rangle-|11\rangle-|22\rangle+|33\rangle,|00\rangle-|11\rangle+|22\rangle-|33\rangle);\nonumber
\\
&(|00\rangle-|11\rangle-i|22\rangle-i|33\rangle,|00\rangle-|11\rangle+i|22\rangle+i|33\rangle,\nonumber\\
&|00\rangle+|11\rangle+i|22\rangle-i|33\rangle,|00\rangle+|11\rangle-i|22\rangle+i|33\rangle)
; \nonumber \\
&\cdots\nonumber\}
\end{eqnarray}
\begin{eqnarray}
&\{(|01\rangle+|12\rangle+|23\rangle+|30\rangle,|01\rangle+|12\rangle-|23\rangle-|30\rangle,\nonumber\\
&|01\rangle-|12\rangle-|23\rangle+|30\rangle,|01\rangle-|12\rangle+|23\rangle-|30\rangle);\nonumber\\
&(|01\rangle-|12\rangle-i|23\rangle-i|30\rangle,|01\rangle-|12\rangle+i|23\rangle+i|30\rangle,\nonumber\\
&|01\rangle+|12\rangle+i|23\rangle-i|30\rangle,|01\rangle+|12\rangle-i|23\rangle+i|30\rangle);\nonumber\\
&\cdots \nonumber\}
\end{eqnarray}
\begin{eqnarray}
&\{(|02\rangle+|13\rangle+|20\rangle+|31\rangle,|02\rangle+|13\rangle-|20\rangle-|31\rangle,\nonumber\\
&|02\rangle-|13\rangle-|20\rangle+|31\rangle,|02\rangle-|13\rangle+|20\rangle-|31\rangle);\cdots\nonumber\\
&\cdots\nonumber\}
\end{eqnarray}
\begin{eqnarray}
&\{(|03\rangle+|10\rangle+|21\rangle+|32\rangle,|03\rangle+|10\rangle-|21\rangle-|32\rangle,
\nonumber\\
&|03\rangle-|10\rangle-|21\rangle+|32\rangle,|03\rangle-|10\rangle+|21\rangle-|32\rangle);
\cdots\nonumber\\
&\cdots\},
\end{eqnarray}
where, for the sake of brevity, we omitted the normalization
factor ($1/2$). Within each set, the four bases are mutually
unbiased, as in (\ref{quartrits}), while the vectors of the bases
from different sets are orthogonal.

As a conclusion one found that the two related concepts of mutual
unbiasedness and maximal entanglement derive from the study of
lifts of the base field $\mathcal{Z}_p$ to Galois fields of prime
characteristic $p>2$ (in odd dimension), or of lifts of the base
ring $\mathcal{Z}_4$ to Galois rings of characteristic $4$ (in
even dimension). One wonders if lifts to more general algebraic
structures would play a role in the study of non maximal
entanglement. We have first in mind the nearfields also useful for
deriving efficient classical codes and which have a strong
underlying geometry.

%
%
%
%


\end{document}